\documentclass[preprint,natbib]{llncs}
\usepackage[utf8]{inputenc}
\usepackage[T1]{fontenc}
\usepackage[breaklinks=true,
            colorlinks=true,
            citecolor=black,
            linkcolor=black, 
            urlcolor=black]{hyperref}

\usepackage{multirow}

\usepackage{charter}
\usepackage{ascii}
\usepackage{tikz}

\usetikzlibrary{decorations.pathmorphing}
\usetikzlibrary{decorations.pathreplacing}
\usepgflibrary{shapes}
\usepgflibrary{arrows}
\usetikzlibrary{patterns}

\usepackage{graphicx,xcolor,listings,amsmath,amssymb}

\def\initlinks{%
  \newwrite\lns%
  \begingroup%
    \catcode`\%=12%
    \immediate\openout\lns=\jobname.lns%
    \immediate\write\lns{\noexpand\newcounter{lnsc}}%
  \endgroup%
}

\def\xxurl#1{
  \begingroup%
  \def\notilda{~}%
  \def~{\~{}}%
  \textit{\scriptsize\href{#1}{#1}}%
  \def~{\notilda}%
  \endgroup%
}

\def\deflink#1#2#3{%
  \immediate\write\lns{%
    \noexpand\refstepcounter{lnsc}%
    \noexpand\noindent{%
      \noexpand\label{link:#1}%
      {\noexpand\small(\noexpand\arabic{lnsc})\noexpand~{#2}}%
    }\noexpand\newline%
  }%
  \begingroup%
  \def\notilda{~}%
  \def~{\noexpand~}
    \immediate\write\lns{%
      \noexpand\indent{\noexpand\xxurl{#3}}\noexpand\newline%
    }%
  \def~{\notilda}%
  \endgroup%
}

\newcommand\link[1]{%
  (\ref{link:#1})%
}

\newcommand\inlinelink[3]{%
  \deflink{#1}{#2}{#3}%
  \link{#1}%
}

\newcommand\insertlinks{%
  \immediate\write\lns{\string }%
  \immediate\closeout\lns%
  \section*{Links}
  \input{\jobname.lns}
}

\initlinks

\begin{document}

\newcommand\ocaml{Objective Caml}
\newcommand\OCAMIL{OCamil}
\newcommand\sml{Standard ML}
\newcommand{\OJACARE}{{O'Jacar\'e}}
\newcommand{\OJACARED}{{O'Jacar\'e.net}}
\newcommand{\DOTNET}{.NET}
\newcommand{\CSHARP}{C\#}
\newcommand{\stopcopy}{Stop\&Copy}

\lstnewenvironment{locaml}[2]{%
  \lstset{basicstyle=\scriptsize,numbers=left,captionpos=b,%
    language=Caml,%
    morekeywords={module, object, sig, struct, %
      val, external, class, functor, open%
    },label={lst:#1},caption={#2}}}{}
\lstnewenvironment{ljs}[2]{%
  \lstset{basicstyle=\scriptsize,numbers=left,%
    captionpos=b,language=JS,label={lst:#1},caption={#2}}}{}
\lstnewenvironment{lhtml}[2]{%
  \lstset{basicstyle=\scriptsize,numbers=left,%
    captionpos=b,language=html,label={lst:#1},caption={#2}}}{}

\newcommand{\ilcodebox}[1]{\mbox{%
    \raisebox{.3ex}{$\ulcorner$}\hspace{-.75ex}%
    {#1}%
    \hspace{-.75ex}\raisebox{.3ex}{$\urcorner$}}}

\newcommand{\ilocaml}[1]{%
  \ilcodebox{\lstinline[language=Caml,%
    morekeywords={%
      module, object, sig, struct,%
      val, external, class, functor, open}]{#1}}}
\newcommand{\iljs}[1]{\ilcodebox{\lstinline[language=JS]!#1!}}
\newcommand{\ilhtml}[1]{\ilcodebox{\lstinline[language=html]!#1!}}

\newenvironment{arevoir}{%
  \color{blue!30!black}%
  \textbf{[À revoir]} %
}{%
}
\newenvironment{arefaire}{%
  \color{red!30!black}%
  \textbf{[À refaire]} %
}{%
}
\newenvironment{NOTE}[1]{%
  \color{yellow!50!black}%
  \textbf{NOTE (#1) :} %
}{%
}

\newcommand{\OC}{O'Caml}
\newcommand{\OComplete}{Objective Caml}

\newcommand{\ang}[1]{\em{#1}}
\newcommand{\lat}[1]{\em{#1}}
\newcommand{\forget}[1]{}

\newcommand{\OCFMC}{OC4MC}
\newcommand{\OCFMComplete}{Objective Caml for Multicore Architectures}

\newcommand{\ET}{Esterel Technologies}
\newcommand{\Scade}{SCADE SUITE 6$^{\mathtt{TM}}$}




\title{{\OCFMC : \OCFMComplete}}
\titlerunning{{\OCFMC}}

\institute{
  \'Equipe APR - D\'epartement CALSCI \\
  Laboratoire d'Informatique de Paris 6 (CNRS UMR 7606) \\
  Universit\'e Pierre et Marie Curie (Paris 6) \\ 
  4 place Jussieu, 75005 Paris, France\\
  \email{\{Mathias.Bourgoin,Benjamin.Canou\}@lip6.fr}\\
  \email{\{Emmanuel.Chailloux,Adrien.Jonquet,Philippe.Wang\}@lip6.fr}\\[10pt]
}

\author{Mathias Bourgoin, Benjamin Canou, Emmanuel Chailloux, Adrien Jonquet and Philippe Wang}
%

\maketitle

\begin{abstract}
  Objective Caml is a famous dialect of the ML family languages. It is
  well-known for  its performance as a  compiled programming language,
  notably  thanks to  its incremental  generational  automatic memory
  collection. However,  for historical  reasons, the latter  was built
  for  monocore processors.   One consequence  is the  runtime library
  assumes there  is effectively no more  than one thread  running at a
  time,  which allows many  optimisations for  monocore architectures:
  very few thread mutexes are sufficient to prevent more than a single
  thread to run  at a time.  This makes  memory allocation and 
  collection quite easier. The way  it was built makes it not possible
  to take  advantage of now  widespread multicore  CPU architectures.
  This  paper  presents  our  feedback on  removing  Objective  Caml's
  garbage  collector  and  designing a  ``Stop-The-World  Stop\&Copy''
  garbage collector to permit threads to take advantage of multicore
  architectures.

  {\keywordname} Objective Caml, Garbage Collector, Parallel Threads, Multicore  
\end{abstract}





\section{Introduction}
%
%









History shows Caml dialects are generally designed 
for monoprocessor architectures. This is actual for Caml (the ancestor),
Caml-Light (on the Zinc machine), Caml-Special-Light and {\OComplete}
(\OC)\inlinelink{Caml}{The Caml Language}{http://caml.inria.fr}\footnote{We use the notation ($n$) to reference external links; see section Links at the end of the document for full URLs.}. 
For the last 25 years, the gain had been more important by
optimizing the sequential runtime libraries or compiler schemes than by modifying
them to target
multiprocessors\inlinelink{GC}{A brief history of Caml}{http://www.pps.jussieu.fr/\%7Ecousinea/Caml/caml\%5Fhistory.html}.
One
of the main success factors of {\OC} was its efficient implementation
from Inria~\cite{ocaml-refman:2008}.\\

 
Currently, {\OC}'s implementation of threads is only a way to express
concurrent algorithms in the language since threads cannot actually be
executed in parallel.  But the recent rise of cheap multicore
architectures -- an average machine can run two or more threads in
parallel -- has created the need to use them.  The {\OC} community
would like concurrent programs to compute faster on these
architectures, as it is already the case for many other languages.\\


One  way to achieve this goal is to add
POSIX-like threads, which can actually run in parallel, to {\ocaml}.
Since the current runtime library has not ben designed with this model in
mind, we need to provide a compatible alternative runtime library, in
particular a new garbage collector, a new allocator and a new thread
library.\\


The idea of modifying {\OC}'s runtime library has already been used to
certify safety-critical software development tools by the French
company {\ET} for its new {\Scade} qualifiable code generator (KCG) on
{\OC} \cite{ICFP2009}.
An {\OC} program such as KCG uses two kinds of
library code: the {\OC} \emph{standard library}, written mainly in {\OC},
and the \emph{runtime library}, written in C and assembly language.
Both are shipped with the {\OC} compiler and linked
with the final executable. The difficulty of specifying and testing
such low-level library code led to adapt and simplify it.
The bulk of the modifications of the \emph{runtime library} was to
remove unessential features according to the coding standard of KCG.
Most of the work consisted in simplifying
the efficient but complex memory management subsystem. {\ET} successfully
replaced it by a plain {\stopcopy} collector with a reasonable loss of
performance~\cite{PADL2008}.

One difficulty of this replacement
was due to the tight coupling of the garbage collector 
with the {\OC} compiler
(in-memory representation of values and entry points of the memory
manager).\\


In this paper we propose to follow this way to exchange the garbage collection runtime
library and thread library to be compliant with POSIX threads that
allow to use parallel threads in a shared-memory concurrency model.
One main constraint is to modify the least possible the {\OC} code
generator and to focus modifications only for the runtime
library. For that, we design a simple garbage collector with a unique
copying, stop-the-world, compacting algorithm.  In this case, the
garbage collector
is sequential and threads can be parallel with a synchronized
mechanism when a garbage collector is called. This step allows to emphasize improved
performances for real parallel programs in {\OC}.
Once the garbage collector interface and thread
library are defined using this assumption, it is possible to
implement other garbage collector algorithms\cite{13}\inlinelink{JonesGC}%
{Richard Jones' Garbage Collection Page}%
{http://www.cs.kent.ac.uk/people/staff/rej/gc.html}.\\

We have completely experimented this way, and the modifications of the Inria distribution are available as a patch called {\OCFMC}\inlinelink{distrib}{{\OCFMC} distribution}
  {http://www.ortsa.com:8480/ocmc/web/}.\\



The rest of this paper is organized as follows.  Section 2 describes
the main features of {\OC}'s runtime library, mainly those which will
change. Section 3 explains how to unplug the original garbage
collector in a sequential world by refering to the {\ET}
experiment. Section 4 shows how to make the runtime library reentrant
and interface its core with its thread library. Section 5 describes
the synchronization mechanism of our garbage collector. Section 6
details the implementation of our garbage collector algorithm. Section
7 presents some {\OC} benchmarks for multicore and comments results.
Section 8 discusses related works while section 9 outlines our future
work.






\section{{\OC}'s runtime  library} 
\label{section:runtimeinria}
%



{\OC}'s  high performance  is partially  due to  its  runtime library,
which  is written  in C,  plus a  per-architecture assembly  file that
allows {\OC}  calls and {C} calls  to live together.   Its original garbage collector
allows a very fast allocation.

\subsection{Garbage Collection}

{\OC} has a two-generation garbage collector, derived from~\cite{158611}. To allocate
a value in the young generation heap, there are two possibilities:
whether there is enough space, in which case a pointer is decremented
by the size of the allocation, or there is not enough space, in which
case the Stop\&Copy garbage collector is triggered.  The Stop\&Copy part of {\OC}'s
garbage collection consists of copying the useful values of the young heap to the old
heap.  The latter is cleaned with an incremental Mark\&Sweep\&Compact
algorithm, which consists of marking the values to sweep the useless
ones, and sometimes compact the heap to take back the empty wholes
left after value sweeping.

Thence {\OC} provides a  particularly efficient allocation, with a Stop\&Copy
part that  is fast because the young  heap is small. We  also have the
possibility  to   program  acceptable  user  interfaces   as  the  old
generation is cleaned  incrementally so that we should  never wait too
long for the garbage collection.


\subsection{Foreign function interface}
For each supported architecture, there is an assembly interface to allow
C function calls and {\OC} function calls to live together. Those files 
also contain fast memory allocation code and exception mechanism code. 
One should not ignore those files when modifying the runtime library
design.

\subsection{Thread library}

{\OC} supports concurrency through:
\begin{itemize}
\item A POSIX threads-like low-level shared memory model, including
  mutexes and conditions.
\item A higher level model based on Concurrent ML~\cite{cml},
  implemented over the low-level thread library.
\end{itemize}

We focus on the low-level thread library, in particular its
implementation and interaction with the runtime library.

\subsection{Thread library implementation}

\indent\indent\indent\indent\emph{or how threads are scheduled to run sequentially}

\subsubsection{POSIX model threads, the simple case}
Current  thread library is  useful to  write concurrent  programs. It
provides  a  set  of  functions  which  match  POSIX  thread  library
functions  for the  C  language. But  will  not allow  threads to  run
simultaneously.   Indeed, they  share a  mutex that  prevents parallel
memory allocations.  Then to start  the Stop\&Copy garbage collection, there is only a
single thread to  stop. As functional programs tend  to allocate a lot
(of small values), it is generally reliable to schedule the threads on
allocation, which  is the mechanism  that is used in  {\OC}.  However,
scheduling  at each  and  every allocation  may  cost too  much, so  a
tick-counting  thread is  launched  to measure  the  time the  current
thread has been running.

\subsubsection{Blocking operation case}
Operations such as I/O operations (e.g. listen on a socket) or locking
operations (e.g.   \texttt{Mutex.lock}) may block a thread  for a long
time. During this,  the thread cannot access the heap  anyway as it is
blocked,  so it  should be  safe  to allow  other threads  to run.   A
mechanism allows to  declare such operations (``enter/leave blocking section'') so that  when they occur,
the scheduler will  detach the thread during the  time of its blocking
operation and  allow another thread to  run. At the end  of a blocking
operation, the thread is attached back to the scheduler.

\subsubsection{Non allocating threads}
A thread that never allocates memory, without ever invoking a blocking
operation, may  prevent the  scheduler to trigger,  and may  block the
whole  program. It  is  assumed that  such  programs do  not occur  in
practice, and  if it ever might  occur then the  programmer should use
\texttt{Thread.yield}.

\section{Replacing {\OC}'s garbage collector: The Esterel Technologies experiment}

Civil  avionics  software  certification authorities  assess  standard
software    engineering    rules    for    safety-critical    software
development. Such software dysfunction may have lethal consequences to
their  users  (flight commands,  railway  traffic  lights, etc).   The
DO-178B  standard  defines all  the  constraints  ruling the  aircraft
software  development.   Code  development  as  it  is  recognized  by
certification authorities follows the  traditional V-Model dear to the
software engineering industry. Traceability during each step of the 
development process is mandatory. 


For   Scade   6,   Esterel   Technologies   decided   to   use   {\OC}
\cite{PADL2008},  which is  very well  suited for
writing compilers, but quite outstanding in a very conservative domain
such as civil  avionics, even though DO-178B encourage  the use of the
best language for a given project. Classical language in this field is
C, or  subsets of C++ or  Ada.  Taking a new  path meant demonstrating
the compatibility  between DO-178B and  {\OC}, by showing  the process
was  under control  (e.g.  generated  code  is as  expected, runtime
library is  predictable).  To do so, Esterel  Technologies decided not
to use the object layer  nor experimental features of {\OC}. While the
standard library was welcome as  fully documented and unit tested, the
runtime  library was  not usable  as such.  Indeed,  to make  it more
understandable, it was partially rewritten, in particular the garbage collection.  The
incremental two-generation garbage collector
was  much too  complex to  certify, and  so was  replaced by  a simple
one-generation  Stop\&Copy garbage collector. This  allowed to  dramatically decrease
the number of lines of code and to make it easy to document (125 lines
of C instead of 1200, for the garbage collector).





This experiment showed it was  feasible to replace some relatively big
parts  of the  runtime  library. We  supposed giving  parallel-capable
concurrent   threads  to   {\OC}  was   feasible  as   well.   Section
\ref{section:reentrance} details this issue.



\subsubsection{How to replace the garbage collector}
Basically, to replace {\OC}'s garbage collector, the process consists of the following steps
\begin{enumerate}
\item regrouping the garbage collector global variables (essentially heap pointers),
\item providing get and set functions over them and use them,
\item and replacing the allocation functions and the collection code.
\end{enumerate}
However, this  works only  if the new  garbage collector still  prevents simultaneous
threads.  Indeed, some global  variables used  in the  runtime library
makes  parallel threads  unsafe, as  they may  use the  same temporary
variable simultaneously.
We address this issue in the next sections.

\section{Runtime library reentrance and parallel threads} 
\label{section:reentrance} 

In the current runtime library source code, threading primitives are
already separated from core functionality. However, the core assumes
that threading primitives allow only one thread to run at a
time. In fact, this source code separation has only been made to
provide several implementations of threads (over POSIX, Win32, etc.),
but does not permit to change the threading model. In this section, we
exhibit the main problems preventing the runtime reentrance and how we
solved them.

\subsubsection{Execution context}
A first problem is the program context being stored in global
variables in the core module. For this to work in presence of threads,
current threading module implementations use
\begin{itemize}
\item a global lock preventing threads to run in parallel,
\item and a context save/restore mechanism in per-thread context
  objects when switching between threads.
\end{itemize}

Our solution for threads to access their execution contexts in
parallel is to leave the threading module handle the context.  Global
variables accesses in the core module are replaced by calls to context
accessors from the threading module.

\subsubsection{Global variables}
Along with the execution context, other global variables exist. For
these, we have to distinguish between three kinds of use:
\begin{enumerate}
\item Temporary (i.e. that does not need to be saved on thread switch)
  thread local data is stored in global variables: we used either the
  same solution as the one for the execution context or more
  lightweight solution like adding function parameters
\item Shared data: we use a global lock mechanism provided by the
  threading module
\item Performance-critical shared data: memory management structures
  for which we had to implement a new parallel-compliant memory
  manager.
\end{enumerate}

\subsubsection{Memory management structures}
There are three main memory management structures.
\begin{enumerate}
\item Heaps which depend on the garbage collection algorithm we will
  use and will be described later
\item Global roots: we used global locks
\item Local roots which are local to threads and contain pointers to
  {\OC} stack segment boundaries and to {\OC} values in C stack
  segments. The local roots are stored in the per-thread structure and
  as we described earlier can be updated concurrently (foreign function interface is
  untouched)
  However, the garbage collector must have access to all roots
  (globals and locals) which can't be writable by other threads during
  a collection. We had to implement a synchronisation mechanism to be
  sure, no thread is able to update its roots during a
  collection. This mechanism will be detailed in the following
  section.
\end{enumerate}

\subsubsection{Interface between the core and threads} To sum up, the core and
threading modules are now interfaced as follows:
\begin{itemize}
\item The threading module defines a way to register thread-local
  data. When the core accesses such data, it must pass through the
  associated accessors. In our implementation, these accessors perform
  direct accesses to per-thread structures. A sequential
  implementation (like OCaml's current one) could simply use global
  variables.
\item For global data, the threading module defines a global lock
  mechanism. In our parallel implementation, we use a mutex lock.
\item The mechanism is generic, however, for performance reasons, we
  provide a way to redefine manually some critical functions, possibly
  in assembly.
\end{itemize}

\section{A garbage collector for parallel threads} 
\label{section:gcforparallelthreads}
%
%




\subsection{Stopping the World}

To run a collection, the garbage collector has to know the exact
state of all  memory management structures (including the local
roots).  The heap has to be inaccessible for every other
thread. Before a collection, the collector has to stop every running
thread, namely: stop the world. Moreover, this mechanism has to stop
the threads while their local roots are updated and exact. One of our
main constraints is to keep the O'Caml compiler unchanged. However,
due to the current compiler, the only moment where the local roots of
a thread are exacts is at an allocation. Thus, there is no other
choice than to stop the threads during an allocation. Besides, this is
the way the current sequential implementation behaves as the collector
can only be triggered after a failed allocation while every thread is
stopped (every thread being stopped during an allocation by the
scheduler).
  
\begin{figure*}
  \centering
  \label{fig:stop_the_world}
  
  \begin{tikzpicture}[decoration={snake, amplitude=0.05cm, segment length=0.3cm}]
    \draw (4, -0.3) -- (4, 1.3) ;
    \draw (5, -0.3) -- (5, 1.3) ;
    \draw (7, -0.3) -- (7, 1.3) ;
    
    \node at (0.9, 0.0) [anchor=east] {3} ;
    \node at (2.9, 0.5) [anchor=east] {2} ;
    \node at (1.9, 1.0) [anchor=east] {1} ;
    
    \node at (2.8, 1.3) [anchor=south] {(a)} ;
    \node at (4.5, 1.3) [anchor=south] {(b)} ;
    \node at (6.0, 1.3) [anchor=south] {(c)} ;
    \node at (8.5, 1.3) [anchor=south] {(d)} ;
    
    \draw [dashed] (1.2,0.0) -- (10, 0.0) ;
    \draw [line width=0.1cm, triangle 90 cap reversed-] (1,0.0) -- (4.5, 0.0) ;
    \draw (3.8, 0.0) [fill=white] circle  (.08cm) ;
    \draw (1.7, 0.0) [fill=white] circle  (.08cm) ;
    \draw (2.3, 0.0) [fill=white] circle  (.08cm) ;
    \draw (2.6, 0.0) [fill=white] circle  (.08cm) ;
    
    \draw (4.5, 0) [fill=white!50!black]  circle  (.08cm) ;
    \draw [line width=0.1cm, -triangle 90 cap] (7,0.0) -- (10, 0.0) ;
    
    \draw (7.0, 0.0) [fill=white] circle  (.08cm) ;
    \draw (7.8, 0.0) [fill=white] circle  (.08cm) ;
    \draw (8.4, 0.0) [fill=white] circle  (.08cm) ;
    \draw (9.5, 0.0) [fill=white] circle  (.08cm) ;

    \draw [dashed] (3.2,0.5) -- (10, 0.5) ;
    \draw [line width=0.1cm, triangle 90 cap reversed-] (3,0.5) -- (4, 0.5) ;
    \draw (3.5, 0.5) [fill=white] circle  (.08cm) ;
    \draw (3.3, 0.5) [fill=white] circle  (.08cm) ;
    \node at (4.0, 0.5) [draw=black, fill=white,, inner sep=.06cm, 
    starburst points=7,
    starburst point height=.1cm, starburst]  {} ;
    \draw [line width=0.1cm, decorate] (5,0.5) -- node [fill=white, midway] {\small GC} (7, 0.5) ;
    \draw [line width=0.1cm, -triangle 90 cap] (7,0.5) -- (10, 0.5) ;

    \draw (7.0, 0.5) [fill=white] circle  (.08cm) ;
    \draw (7.3, 0.5) [fill=white] circle  (.08cm) ;
    \draw (7.7, 0.5) [fill=white] circle  (.08cm) ;
    \draw (8.0, 0.5) [fill=white] circle  (.08cm) ;
    \draw (8.7, 0.5) [fill=white] circle  (.08cm) ;
    \draw (9.2, 0.5) [fill=white] circle  (.08cm) ;
    \draw (9.4, 0.5) [fill=white] circle  (.08cm) ;
    \draw (9.7, 0.5) [fill=white] circle  (.08cm) ;

    \draw [dashed] (2.2,1.0) -- (10, 1.0) ;
    \draw [line width=0.1cm, triangle 90 cap reversed-] (2,1.0) -- (5, 1.0) ;
    \draw (2.2, 1.0) [fill=white] circle  (.08cm) ;
    \draw (3.3, 1.0) [fill=white] circle  (.08cm) ;
    \draw (3.7, 1.0) [fill=white] circle  (.08cm) ;
    \draw (5, 1.0) [fill=white!50!black]  circle  (.08cm) ;
    \draw [line width=0.1cm, -triangle 90 cap] (7,1.0) -- (10, 1.0) ;

    \draw (7.0, 1.0) [fill=white] circle  (.08cm) ;
    \draw (8.7, 1.0) [fill=white] circle  (.08cm) ;
  \end{tikzpicture}

  \caption{Stop \& Copy}
  \vspace{1ex}

  \begin{tikzpicture}[decoration={snake, amplitude=0.05cm, segment length=0.3cm}]   
    \draw [line width=0.1cm, -triangle 90 cap] (0,0) -- +(0.9cm,0) ;
    \node at (1, 0) [anchor=west] { \small : Running thread } ;

    \draw [line width=0.1cm, decorate, -triangle 90 cap] (4,0) -- +(0.9cm,0) ;
    \node at (5, 0) [anchor=west] { \small : Thread doing GC } ;

    \draw [dashed] (8,0) -- +(0.9cm, 0) ;
    \node at (9, 0) [anchor=west] { \small : Paused thread{\color{white}g} } ;
  \end{tikzpicture}

  \begin{tikzpicture}[decoration={snake, amplitude=0.05cm, segment length=0.3cm}]
    \node at (0, 0) [draw=black, fill=white,, inner sep=.06cm, 
                          starburst points=7,
                          starburst point height=.1cm, starburst]  {} ;
    \node at (.3, 0) [anchor=west] { \small : Failed allocation{\color{white}p}} ;

    \draw (3.5, 0) [fill=white] circle  (.08cm) ;
    \node at (3.8, 0) [anchor=west] { \small : Successful allocation{\color{white}p} } ;
    \draw (7.3, 0) [fill=black!50!white] circle  (.08cm) ;
    \node at (7.6, 0) [anchor=west] { \small : Suspended allocation} ;
  \end{tikzpicture}
  
\end{figure*}
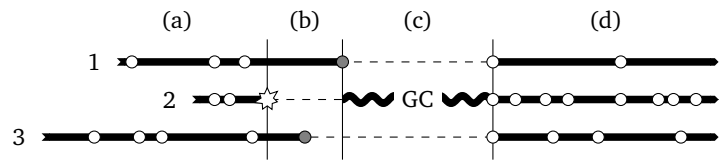

Fig.~\ref{fig:stop_the_world} describes the stop the world mechanism
implemented, with three threads.
\begin{itemize} 
\item[(a)] Each thread may allocate until thread 2 fails an
  allocation, which means the need of a collection.
\item[(b)] Then, every other running thread will stop on the next
  allocation (ensuring the correctness of its local roots).
\item[(c)] Every thread stopped, the garbage collector runs.
\item[(d)] Each thread resumes its pre-collection behaviour starting
  by the allocation which made it stop.
\end{itemize}
As in the original implementation we use the enter/leave blocking
section mechanism (described section 2.4) to allow blocking operation
without preventing the stop of the world. Stopping the
world allows the implementation of a sequential garbage collector.

\subsection{Interface between thread library and garbage collector}

To use various garbage collectors with different thread library
implementations, we had to define clearly the interactions between the
two.

As the core runtime library, the garbage collector uses accessors
provided by the thread library. We also added primitives to iterate
over threads so that the garbage collector does not have to know how
thread-local data are stored.

On its side, the garbage collector defines primitives that has to be
used by the thread library implementation to ensure that the system is
in a correct state. For example, the garbage collector has to be
notified when a thread has been paused. Indeed, a garbage collector
must not wait for paused threads when stopping the world, and has to
prevent them to be resumed before the collection has ended.

\section{Our garbage collector}
\begin{figure}[hbt]
  \centering

\begin{tikzpicture}
  \fill [fill=black!20!white] (9, -0.5) rectangle (10, 0.5) ;
  \draw (7, -0.5) rectangle (10, 0.5) ;
  \draw (9, -0.5) -- (9, 0.5) ;

  \node (SE) at (7, 1) { \small End } ;
  \draw [->] (SE.south) -- (7, .5) ;
  
  \node (SP) at (9, 1) { \small Pointer } ;
  \draw [->] (SP.south) -- (9, .5) ;
  
  \node (SS) at (10, 1) { \small Start } ;
  \draw [->] (SS.south) -- (10, .5) ;

  \begin{scope}[xshift=7cm]
    \fill [fill=black!20!white] (8, -0.5) rectangle (10, 0.5) ;
    \draw (7, -0.5) rectangle (10, 0.5) ;
    \draw (9, -0.5) -- (9, 0.5) ;
    \draw (8, -0.5) -- (8, 0.5) ;
    \draw [decorate, decoration={brace, mirror, amplitude=0.2cm}] (8, -0.7) -- (9, -0.7);
    \node [anchor=north] at (8.5, -1) {allocated block} ;

    \node (SE) at (7, 1) { \small End } ;
    \draw [->] (SE.south) -- (7, .5) ;
    
    \node (SP) at (8, 1) { \small Pointer } ;
    \draw [->] (SP.south) -- (8, .5) ;
  
    \node (SS) at (10, 1) { \small Start } ;
    \draw [->] (SS.south) -- (10, .5) ;
  \end{scope}

  \draw [->] (10.2, 0) -- node [fill=white, above] {Pointer = Pointer - size} (13.8, 0) ;

\end{tikzpicture}

  \caption{Allocation in OCaml's small heap}
  \vspace{1ex}
  
  \begin{tikzpicture}
    \node (OS) [shape=rectangle, inner sep=1ex, draw, fill=black!20!white] {} ;
    \node (O) [right of=OS] {\small Used space} ;
    \node (FS) [right of=O, xshift=1cm, shape=rectangle, inner sep=1ex, draw] {} ;
    \node (F) [right of=FS] {\small Free space} ;
  \end{tikzpicture}
  \label{fig:caml_heap}
\end{figure}
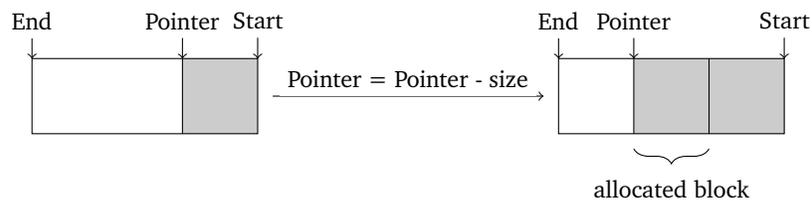

For a language as {\OC}, it is very important to have a very low cost
allocation mechanism for small objects. {\OC}'s current allocator
dedicates a small (young) heap to small objects. This heap is a
contiguous memory segment with a cursor indicating the end of the used
zone, as shown in Fig.~\ref{fig:caml_heap}.

Thus, allocating is simply decreasing the cursor, testing if it has not
crossed the limit, and returning the new cursor as the pointer to the
newly allocated block.  Of course, in presence of concurrent access to
the heap variables, this is not possible. Moreover, adding a lock
mechanism around the allocation would be too costly.

We will now describe our (fairly simple) memory management solution,
which provides fast allocation for small objects, but allows several
threads to allocate at the same time.

We first describe the structure of our heap, then we describe the
garbage collection mechanism. We first describe the simple version,
called \textit{full} collection, then a more complex variation adding
\textit{partial} collection cycles.

\subsection{Heap structures}

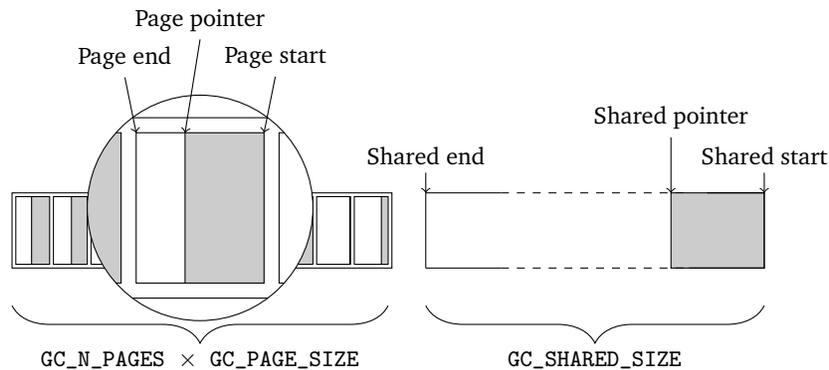
\begin{figure}[hbt]
  \centering

\begin{tikzpicture}
  \pgfmathrandominteger{\b}{0}{100};
  \pgfmathrandominteger{\c}{0}{27};
  \draw (0, -0.5) rectangle (5.05, 0.5) ;
  \foreach \i in {0.05, 0.55, ..., 4.55} {
    \pgfmathrandominteger{\a}{0}{45};

    \fill [black!20!white] (\i+\a/100, -0.45) rectangle (\i+ 0.45, 0.45) ;
    \draw (\i+\a/100, -0.45) -- (\i+\a/100, 0.45) ;
    \draw (\i, -0.45) rectangle (\i+0.45, 0.45) ;

  } ;

  \begin{scope}[yshift=0.3cm]
    \fill [white] (2.5, 0) circle (1.5cm) ;
    \begin{scope}
      \clip (2.5, 0) circle (1.5cm) ;
      \draw (0,-1.2) -- (5, -1.2) ;
      \draw (0,1.2) -- (5, 1.2) ;
      \fill [black!20!white] (2.30,-1) rectangle (3.35,1) ;
      \draw (1.65,-1) rectangle (3.35,1) ;
      \draw [fill=black!20!white] (0,-1) rectangle (1.45,1) ;
      \draw (3.55,-1) rectangle (5,1) ;
      \draw (2.3,-1) -- (2.3,1) ;
    \end{scope}
    \draw (2.5, 0) circle (1.5cm) ;
    
    \node (PE) at (1.5, 2) { \small Page end } ;
    \draw [->] (PE.south) -- (1.65, 1) ;
    
    \node (PP) at (2.5, 2.5) { \small Page pointer } ;
    \draw [->] (PP.south) -- (2.3, 1) ;
    
    \node (PS) at (3.5, 2) { \small Page start } ;
    \draw [->] (PS.south) -- (3.35, 1) ;
  \end{scope}

  \draw [decorate, decoration={brace, mirror, amplitude=.5cm}] (0, -1) -- (5, -1);
  \node [anchor=north] at (2.5, -1.5) {\footnotesize \tt GC\_N\_PAGES $\times$ GC\_PAGE\_SIZE} ;

  \draw (5.5, -0.5) -- (5.5, .5) -- (6.5,.5) ;
  \draw [dashed] (6.5, .5) -- (9, .5) ;
  \draw [fill=black!20!white] (9-\b/100, -0.5) rectangle (10, 0.5) ;
  \draw (9+\b/100, .5) -- (10, .5) -- (10, -0.5) -- (9, -0.5) ;
  \draw [dashed] (6.5, -0.5) -- (9, -0.5) ;
  \draw (6.5, -0.5) -- (5.5, -0.5) ;

  \node (SE) at (5.5, 1) { \small Shared end } ;
  \draw [->] (SE.south) -- (5.5, .5) ;
  
  \node (SP) at (9-\b/100, 1.5) { \small Shared pointer } ;
  \draw [->] (SP.south) -- (9-\b/100, .5) ;
  
  \node (SS) at (10, 1) { \small Shared start } ;
  \draw [->] (SS.south) -- (10, .5) ;

  \draw [decorate, decoration={brace, mirror, amplitude=.5cm}] (5.5, -1) -- (10, -1);
  \node [anchor=north] at (7.75, -1.5) {\footnotesize \tt GC\_SHARED\_SIZE} ;

\end{tikzpicture}

  \caption{Heap structures}
  \vspace{1ex}
  
  \begin{tikzpicture}
    \node (OS) [shape=rectangle, inner sep=1ex, draw, fill=black!20!white] {} ;
    \node (O) [right of=OS] {\small Used space} ;
    \node (FS) [right of=O, xshift=1cm, shape=rectangle, inner sep=1ex, draw] {} ;
    \node (F) [right of=FS] {\small Free space} ;
  \end{tikzpicture}
  \label{fig:heap}
\end{figure}

We shall first present the structure of our heap. A graphical
representation is shown in Fig.~\ref{fig:heap}.

\subsubsection{Local heaps}

Each thread has a small heap using the same cursor mechanism as
{\OC}'s small heap. These small heaps are called pages, in a page
table. Hence, threads can allocate small objects in their own pages
simultaneously safely. When a thread has filled its page, it takes a
new page in the table, in mutual exclusion with others.

\subsubsection{Shared heap}

For bigger objects, a big shared heap is used. Each allocation
requires a lock. Its size varies on demand at every full collection,
as we explain below.

\subsection{Full collection}



This first form of garbage collection cycle is called \textit{full}
since it operates on the whole heap. It is not the default algorithm
but can be enabled by setting the compilation option
\texttt{GC\_ENABLE\_PARTIAL} to \texttt{0}.

\subsubsection{Algorithm}
\label{gc_algo}
When all pages are taken and a thread fails to allocate into one of
its pages, or an allocation into the shared heap fails, then a
collection is triggered.

\begin{figure}[hbt]
  \centering

\begin{tikzpicture}[decoration={snake, amplitude=0.05cm, segment length=0.3cm}]
  \pgfmathrandominteger{\b}{0}{100};
  \pgfmathrandominteger{\c}{0}{27};
  \draw (0, 0) rectangle (5.05, 1) ;
  \foreach \i in {0.05, 0.55, ..., 4.55} {
    \pgfmathrandominteger{\a}{0}{45};

    \fill [black!20!white]
      (\i+\a/100, 0.05)
      .. controls +(0,-0.8) and +(0, 0.8) ..
      (8-\c/100, -1.0) -- (9.33-\b/100, -1.00)
      .. controls +(0,0.8) and +(0, -0.8) ..
      (\i+ 0.45, 0.05) ;

    \fill [black!20!white] (\i+\a/100, 0.05) rectangle (\i+ 0.45, 0.95) ;
    \draw (\i+\a/100, 0.05) -- (\i+\a/100, 0.95) ;
    \draw (\i, 0.05) rectangle (\i+0.45, 0.95) ;
    \draw (\i, -1.95) rectangle (\i+0.45, -1.05) ;
  } ;

  \fill [black!20!white]
  (10, 0.05) .. controls +(0,-0.8) and +(0, 0.8) .. (10, -1.0) 
  --
  (8.7, -1.00) .. controls +(0,0.8) and +(0, -0.8) .. (7.5, 0.05) ;

  \draw (0, -2) rectangle (5.05, -1) ;
  \draw (5.5, 0) -- (5.5, 1) -- (6.5,1) ;
  \draw [dashed] (6.5, 1) -- (7.5, 1) ;
  \draw [fill=black!20!white] (7.5, 0.00) rectangle (10, 1) ;
  \draw (9+\b/100, 1) -- (10, 1) -- (10, 0) -- (9, 0) ;
  \draw [dashed] (6.5, 0) -- (7.5, 0) ;
  \draw (6.5, 0) -- (5.5, 0) ;

  \node (X) at (-1, 0.5) {Before} ;

  \begin{scope}[yshift=-2cm]
  \draw (6.7, 0) -- (6.7, 1) ;
  \draw (6.7, 1) -- (9, 1) ;
  \draw (9+\b/100, 1) -- (10, 1) -- (10, 0) -- (9, 0) ;
  \draw (6.7, 0) -- (9, 0) ;
  \draw [fill=black!20!white] (8-\c/100, 0.00) rectangle (9-\b/200, 1) ;
  \draw [fill=black!20!white] (9-\b/200, 0.00) rectangle (10, 1) ;
  \node (Y) at (-1, 0.5) {After} ;
  \end{scope}
  \draw [line width=0.1cm, decorate, -triangle 90 cap]
  (X.south) -- node [fill=white, midway] {\small Copy} (Y.north) ;
  \node (XXX1) at (8, 0.5) {old shared heap};
  \node (XXX2) at (8.5, -1.5) {new shared heap};
\end{tikzpicture}
  \caption{Full garbage collection cycle}
  
  \label{fig:gc}
\end{figure}
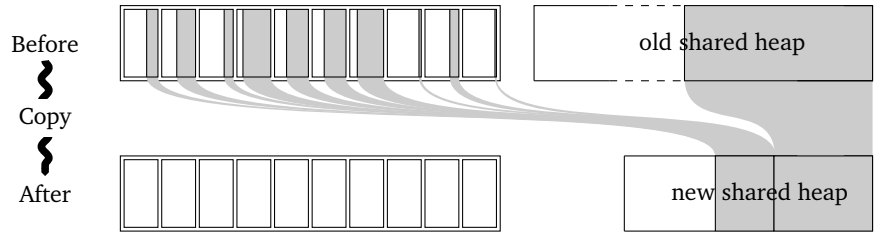

We use a Stop\&Copy algorithm to copy all living values from the pages
and the shared heap into a new shared heap (Figure~\ref{fig:gc}).

Every living heap-allocated value is copied into a new shared heap.
The set of pages is then cleared and each threads gets a new empty
page.  In the end, we have a set of empty pages, and a new shared heap
containing all (and only) living heap-allocated values.

\subsubsection{Heap size}
The number of pages can be set with the \mbox{\texttt{GC\_PAGE\_SIZE}}
environment variable, and their size can be set with the
\mbox{\texttt{GC\_N\_PAGES}} environment variable.
The \mbox{\texttt{GC\_SHARED\_HEAP}} environment variable defines
the size of the shared heap.

While running, the amount of memory needed by the program may
increase. We thus added a mechanism to automatically grow and shrink
the shared heap. For this, the new shared heap allocated during a full
collection is set to the sum of all used space (in pages and shared
heap).

With this solution, if the shared heap is full and entirely alive and
the pages are empty, the new shared heap ends up being already
full. This new size is thus weighted by some constant $k > 1$ (we
chose 1.5 in practice).
 $$size(new\ shared\ heap)=k \times (used\ space(old\ shared\ heap)+used\ space(page\ set))$$

 If the weighted new size is not sufficient to contain the failed
 allocation (i.e. if the programmer tried to allocated more than $k -
 1$ times the size of the old heap) the new size is maximized
 accordingly.

 These precautions are sufficient to prevent triggering the GC in an
 infinite loop, but if the heap shrinks too much, it might be
 triggered too often. To solve this, we also added a minimal,
 configurable size to the heap.




\subsection{Partial collection}

This second form of collection is used complimentarily to the full
collection. It is triggered when an allocation in a page fails and
there is no more page left, but there is enough space in the shared
heap store local values.

The performance of the two versions will be discussed after the
results are presented.

\begin{figure}[hbt]
  \centering

\begin{tikzpicture}[decoration={snake, amplitude=0.05cm, segment length=0.3cm}]
  \pgfmathrandominteger{\b}{0}{100};
  \pgfmathrandominteger{\c}{0}{27};
  \draw (0, 0) rectangle (5.05, 1) ;
  \foreach \i in {0.05, 0.55, ..., 4.55} {
    \pgfmathrandominteger{\a}{0}{45};

    \fill [black!20!white]
      (\i+\a/100, 0.05)
      .. controls +(0,-0.8) and +(0, 0.8) ..
      (8-\c/100, -1.0) -- (9-\b/100, -1.00)
      .. controls +(0,0.8) and +(0, -0.8) ..
      (\i+ 0.45, 0.05) ;

    \fill [black!20!white] (\i+\a/100, 0.05) rectangle (\i+ 0.45, 0.95) ;
    \draw (\i+\a/100, 0.05) -- (\i+\a/100, 0.95) ;
    \draw (\i, 0.05) rectangle (\i+0.45, 0.95) ;
    \draw (\i, -1.95) rectangle (\i+0.45, -1.05) ;
  } ;

  \draw (0, -2) rectangle (5.05, -1) ;
  \draw (5.5, 0) -- (5.5, 1) -- (6.5,1) ;
  \draw [dashed] (6.5, 1) -- (9, 1) ;
  \draw [fill=black!20!white] (9-\b/100, 0.00) rectangle (10, 1) ;
  \draw (9+\b/100, 1) -- (10, 1) -- (10, 0) -- (9, 0) ;
  \draw [dashed] (6.5, 0) -- (9, 0) ;
  \draw (6.5, 0) -- (5.5, 0) ;

  \node (X) at (-1, 0.5) {Before} ;

  \begin{scope}[yshift=-2cm]
  \draw (5.5, 0) -- (5.5, 1) -- (6.5,1) ;
  \draw [dashed] (6.5, 1) -- (9, 1) ;
  \draw (9+\b/100, 1) -- (10, 1) -- (10, 0) -- (9, 0) ;
  \draw [dashed] (6.5, 0) -- (9, 0) ;
  \draw (6.5, 0) -- (5.5, 0) ;
  \draw [fill=black!20!white] (8-\c/100, 0.00) rectangle (9-\b/100, 1) ;
  \draw [fill=black!20!white] (9-\b/100, 0.00) rectangle (10, 1) ;
  \node (Y) at (-1, 0.5) {After} ;
  \end{scope}
  \draw [line width=0.1cm, decorate, -triangle 90 cap]
  (X.south) -- node [fill=white, midway] {\small Copy} (Y.north) ;

  \node (XXX1) at (8, 0.5) {old shared heap};
  \node (XXX2) at (8, -1.5) {(same) old shared heap};
\end{tikzpicture}
  \caption{Partial garbage collection cycle}
  
  \label{fig:gcpart}
\end{figure}
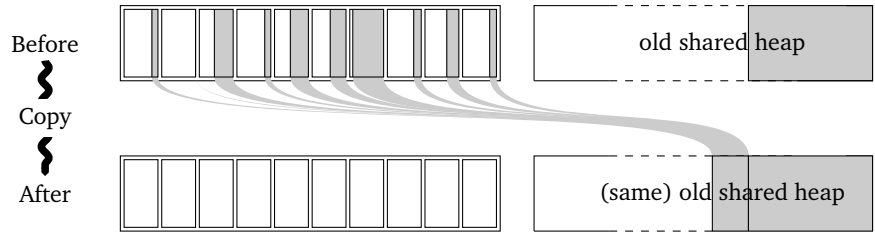

Again, a Stop\&Copy algorithm is used to flush alive values from the
entire page table into the shared heap.  The pages are then
reinitialized and reassigned. This algorithm is graphically
represented by Fig.~\ref{fig:gcpart}.

\subsubsection{Backward pointers}

In a Stop\&Copy algorithm, the values considered alive are the ones
that can be reached from the roots by following pointers in the
from-space. When a pointer leads out of the from-space, it is not
followed. A problem arises when the programmer affects a value in its
page to a mutable cell in the shared heap, creating a \textit{backward
  pointer}, and then forgets this value. In this case, the value is
indeed alive: there exists a path leading to it from the roots, but
this path is not entirely in the from-space. To prevent this kind of
values from being freed on a local collection, we add them explicitely
to the roots on such an affectation.

We use linked lists of memory chunks (instead of singleton pointers
for efficiency) containing backward pointers to alive values. Every
thread has its own list so we don't introduce mutexes. On a partial
collection, we consider them as part of the root set. They are of
course discarded just after the cycle since the values have been
copied to the shared heap.



\subsection{Optimisations and limitations}


For performance reasons, we had to use macros (or inline functions)
and link statically the three parts described above. This does not
hinder the clear separation we have shown at source level, but means
that the garbage collector cannot be changed at run or load
time. Moreover, to achieve good performance, even if a generic version
calling the C function of the garbage collector can be used, writing
an assembly version of the allocator using the garbage collector
structures was necessary to obtain best performance.

\section{Benchmarks}
%


Our performance benchmarks were made on Mac Pro running Ubuntu Linux 9.10
for x86-64bit architecture. The following gives some details about the hardware: 
\begin{itemize}
\item The Mac Pro hosts two quad-core CPUs (Intel Xeon 2.8~GHz, without hyperthreading) which makes a total of 8 cores.
The memory speed is 800~MHz and its capacity is 2$\times$2~GB (4~GB in dual channel),
and the bus speed is double (1.6~GHz).
The number of cores determines the number of possible parallel threads.
The memory speed is mandatory since with multicores, it easily becomes too low 
because it is shared between cores.
\end{itemize}
Our benchmarks were made with the first form of garbage collection
(without partial collection) as described in \ref{gc_algo}.

Since there was very little hope of taking advantage of multicore 
by programing threads in {\OC}, there are very few existing 
benchmarks. Thus we distinguish two types of programs:
\begin{itemize}
\item the classic Caml benchmarks, such as Knuth-Bendix (KB),
which are computed several times within threads,
\item some programs written for the occasion, such as the sieve of
Eratosthenes which is written in two very different paradigms, which
are written with concurrency (and possibly parallelism) in mind.
\end{itemize}
Then, some programs will use a number of concurrent threads 
lesser or proportional to the avaiable number of cores,
and other programs use a great number of concurrent threads.

\subsection{Sequential programs}

{\OC} programs are mainly sequential. We compared our implementation
with {\OC} for sequential programs. Those tests were only testing our
allocator and collector as the runtime is mainly the original {\OC}
one and as the thread library we modified is unused in sequential
programs. 

Our benchmarks consist of the following programs:

 \begin{itemize}
 \item \textbf{KB}: the Knuth-Bendix completion algorithm. This is a
   fully functional program using exceptions intensively to compute
   over terms
    \item \textbf{Nucleic}: floating-point calculations involving trees \cite{Pseudoknot-96}.
    \item \textbf{FFT}: Fast Fourrier Transform, computing floats.
    \item \textbf{QuickSort}: classic (in-place) quick sort algorithm for arrays.
 \end{itemize}

\begin{figure}
  \centering
\begin{tabular}{|l|| c|cccc|}
\hline
 & \textbf{{\OC}} &  \multicolumn{4}{c|}{\textbf{OC4MC}} \\ 
 \# threads & 1TH &  1TH &  2TH &  4TH & 7TH \\\hline
 \textsc{KB} & 4.90s & 9.47s & 14.17s
 & 20.16s & 22.36s\\\hline

\textsc{Nucleic}& 1.20s & 3.98s & 5.29s & 4.99s & 4.31s \\\hline

 \textsc{FFT}& 2.21s & 4.10s & 3.13s & 3.10s & 3.04s\\
\hline
 \textsc{QuickSort}& 9.93s & 9.71s & 5.01s & 2.65s & 1.81s\\
\hline  
\end{tabular}\\
  \caption{Benchmarks for sequential programs}
  \label{fig:bench_seq}
\end{figure}

Those tests allowed us to confirm that our allocator and
garbage collector were less efficient than the original {\OC} ones. As
shown in fig.~\ref{fig:bench_seq}, most programs are slower with OC4MC than with
{\OC} with the exception of programs with a high computation over
allocation ratio.

\subsection{Parallel programs}

We wrote the following specific test programs:

\begin{itemize}
  \item \textbf{sieve}: the sieve of Eratosthenes. It starts with a
    big allocation of a Boolean matrix. Each cell represents an
    integer.  Then each thread removes all multiples of uncomputed
    integers. We ran the tests with integers between 2 and \mbox{300~000}.
    
  \item \textbf{matmult}: a simple matrix multiplication. The main
    loop is parallelized each thread computing its own lines of the
    final matrix. Matrix multiplication naive algorithm has a $O(n^3)$
    complexity which means that the ratio of computation over
    allocation is very high. For our benchmarks we multiplied
    1000$\times$1000 matrixes.
    
  \item \textbf{life}: the classical game of life: a cellular
    automaton.It's an imperative object oriented program. It generates
    a universe (a board) where initialy three cells are alive. Each
    thread manages a section of this universe updating its cells at
    each step of the program. At the end of a step the main thread
    awaits for the other threads to end their calculation before
    allocating a new updated board and discard the old board. For each
    updated cell, dead or alive, a new object is allocated. For our
    benchmarks we limited the universe to a 200$\times$200 board.
  \item \textbf{pi}: a $\pi$ computation. For each point in a square,
    pi tests if the point is included inside the incircle of the
    square. The number of points inside this circle divided by the
    total number of points inside the square gives $\pi/4$. The square
    is divided equally into subsquares, each thread computes a
    subsquare.
\item \textbf{sieve in CML-style}: $n$ successive integers are
  passed through non-prime-number filtering channels.  The first
  filter removes all multiples of 2. For each filter, when the first
  passing number is found (which is a prime number), a new filter
  initiates to filter its multiples. At the end, each created
  filter corresponds to a created thread and to a prime number.
  We ran the tests with integers between 2 and 9000, which creates about 1100 parallel threads. 

\end{itemize}

\begin{figure}
  \centering
\begin{tabular}{|l|| c|ccccc|}
\hline
 & \textbf{{\OC}} &  \multicolumn{5}{c|}{\textbf{OC4MC}} \\ 
 \# threads & 1TH &  1TH &  2TH &  4TH & 7TH & 14TH \\\hline
 \textsc{Sieve} & 68.68s & 68.67s & 38.08s
 & 20.01s & 11.13s & 8.67s
\\
\emph{speedup} & \textbf{1} & \textbf{1} & \textbf{1.80}&
 \textbf{3.43}& \textbf{6.16} &  \textbf{7.92}
   \\\hline
 \textsc{Matmult}& 16.49s & 17.64s & 8.70s & 4.55s & 2.70s & 2.40s\\
\emph{speedup} & \textbf{1.06} & \textbf{1} & \textbf{2.02}&
 \textbf{3.87}& \textbf{6.53} & \textbf{7.35} 
   \\\hline
%
 \textsc{Life}& 13.15s & 16.80s & 12.29s & 10.21s & 10.25s & 9.97s\\
\emph{speedup} & \textbf{1.28} & \textbf{1} & \textbf{1.36}&
 \textbf{1.65}& \textbf{1.64} & \textbf{1.69} \\\hline

 \textsc{Pi}&  22.88s & 22.79s & 11.38s & 5.69s & 3.26s & 2.88s\\
\emph{speedup} & \textbf{0.99} & \textbf{1} & \textbf{2.4}&
 \textbf{4.01}& \textbf{7.0} & \textbf{7.91} \\\hline


  
\end{tabular}\\
  \caption{Benchmarks for little number of threads}
  \label{fig:bench1}
\end{figure}

\begin{figure}
  \centering
  \begin{tabular}{|l||c|c|}
    \hline
     & \textbf{{\OC}} & \textbf{OC4MC} \\
    \hline
    \textsc{Sieve} \textsc{CML-style}& 89s & 59s \\
    \hline
  \end{tabular}
\caption{Benchmarks for great number of threads}
  \label{fig:bench2}
\end{figure}




\subsection{Comparison with {\OC}}

\newcommand{\suppr}[1]{}
\suppr{

Then, we tested parallel programs. Running parallel programs with {\OC} is
often less efficient than running the same program sequentially. It is mainly
due to the thread scheduler of {\OC} which can stop threads a long time,
especially if the program use a lot of critical  sections protected by mutexes
or conditions. Comparing threaded programs with OC4MC and {\OC} is most of the
time irrelevant.  We compared our multithreaded programs running with OC4MC
with the sequential equivalent programs running with {\OC}.

As shown figure~\ref{bench1}, multithreaded programs are running faster than
the sequential ones with OC4MC. Compared with {\OC} the results are depending
of the program we test. For a program with a high computing over allocating
ratio, OC4MC offers a speedup of $N$ for a $N$-core machine. Increasing the
allocation over the computation implies increasing the number of garbage
collection in wich we spend more time than {\OC}. Meanwhile, OC4MC is still
faster than {\OC} for most well parallelized programs. The real weakness
of multithreading comes with programs allocating a lot and making a lot of
memory accesses because having multiple threads running  with shared memory
means dividing the memory bandwidth within the threads leading to a bottleneck
slowing the execution of the program. In such case OC4MC runs programs slower
than {\OC}.
}

For most sequential programs, our alternative runtime library will 
provide predictable yet acceptable loss of performance, because for
sequential programs, our garbage collector is particularly naive
comparing to {\OC}'s. It is also with no surprise that loss of
performance is higher with programs that intensively allocate 
short-life small data. In this case, multithreading may as much allow
a gain of performance as a loss of performance, depending on how
memory is used: multithreading may increase or decrease cache miss.

However, as Fig.~\ref{fig:bench1} shows, OC4MC may also provide speedup.
Indeed, with \emph{sieve} and \emph{matmult} show that the speedup can be close
to the number of cores, while \emph{life} shows that the speedup 
is actual even though limited, probably due to memory speed limitation as
it is a program that allocate a lot of small data.

Fig.\ref{fig:bench2}  shows that  thread intensive  programs  may gain
speedup   thanks  to  switching-cost   limitation,  even   with  heavy
structures.   Plus, we  may  gain performance  with existing  programs
written without parallelism but concurrency in mind. However, we shall
not forget that shared-memory concurrent programming is hard, and adding parallelism
for better performance does not ease the task as computation may actually
run in parallel which makes debugging even harder.

Eventually, with  a relatively simple garbage  collector (which offers
space    for   optimizations    and   improvements),    OC4MC   
brings and shows interesting  performance compared
to {\OC} when  run on multicore architectures, by  allowing threads to
allocate memory (and compute) in parallel.

In~\cite{Harrop}, Jon Harrop compared different versions of the matrix
multiplication program using higher level abstractions such as the
"parallel for" loop. He also tried to use a higher order
implementation of the program which decreased scalability. This has
been corrected in the latest versions of OC4MC. This shows OC4MC
allows the use of functionnal abstractions to allow high level
parallelism without decreasing its performances.

\subsection{Impact of partial collections}

\begin{figure}
  \centering
\begin{tabular}{cc}
\begin{tabular}{|l|| ccccc|}
  \hline
  &   \multicolumn{5}{c|}{\textbf{OC4MC}} \\ 
  \# threads &  1TH&  2TH &  4TH & 7TH & 14TH \\\hline
 \textsc{Sieve} & 68.67s & 38.08s & 20.01s &
 11.13 & 8.67s \\
 \textsc{Sieve partial} & 68.67s & 34.39s &
 17.4s & 10.0s & 8.63\\
\hline 
\hline
 \textsc{Life} & 16.80s & 12.29s & 10.21s &
 10.25 & 9.97s \\
 \textsc{Life partial} & 13.65s & 8.57s &
 7.18s & 7.34s & 7.30\\
\hline



  
\end{tabular}
&

  \begin{tabular}{|l||c|}
    \hline
     & \textbf{OC4MC} \\
    \hline
    \textsc{Sieve} \textsc{CML-style} & 59s \\
    \textsc{Sieve} \textsc{CML-style partial} & 59s \\
    \hline
  \end{tabular}
\end{tabular}
  \caption{Benchmarks with partial collection}
  \label{fig:benchs_partial}
\end{figure}

The perfect case to show better performance when enabling partial
collections is quite clear: a program allocating big data when
starting, then allocating a lot of short-life data in every thread.
\begin{itemize}
\item The drawback of having a faster collection cycle operating over
  less data is straightforward: it has to be triggered more
  often. With a Stop The World technique, this means a greater delay
  for synchronization. If every thread allocates frequently, this
  delay can remain short.
\item When a program allocates big data once and for all, having
  partial collections can importantly reduce collection times.
\item If the data allocated are short-lived, then the amount of space
  copied to the shared heap is minimal and full collections are
  triggered less often.
\end{itemize}

Of course, one could also find worst cases, that is why we provide the
option to use both algorithms.

In practice, on our example set, it seems that enabling partial
collections is most of the time a good idea. An example is shown in Fig.~\ref{fig:benchs_partial}.







\section{Related works}
%
%

%









There have been many attempts to give multicore support for typed
functional languages, whether by adapting runtime libraries or by giving
language extensions for parallel programming, or both.

\cite{runtimehaskell}  presents  runtime  support  for  multicores  in
parallel Haskell.   Parallelism is  explicited by using  a ``\texttt{par}''
combinator, allowing the computation of its first  parameter by a
task (``\texttt{spark}'') manager at anytime while the evaluation continues.
This runtime uses a parallel garbage collector and different optimization techniques
are presented in the paper.

Manticore   \cite{manticore}  is   a  SML-based   functional  language
providing  two  new features:  CML's  thread  mechanism
 (first-class synchronous communication), and  parallel
structures (e.g.  arrays) for data parallelism.  Its two-generation
garbage collector
works with  a per-thread heap  without backward pointers and  a shared
heap. Each thread may collect on its own local heap. If collection fails,
then a global garbage collector is triggered to copy from local heaps to shared heap.

Other approaches have introduced basic parallel computation
mechanisms.  CoThreads
\inlinelink{cothreads}{CoThreads}{http://www.pps.jussieu.fr/~li/software/}
for {\OC} introduces communication between processes which may run on
different processors, with the same interface as {\OC} thread library.
There is currently an attempt to give multicore support for Mlton/SML 
\inlinelink{sml}{Multicore MLton}{http://www.cs.cmu.edu/~spoons/parallel/}, by
implementing low-level threads with additional higher level abstractions. 
F\# (which combines the functional and imperative core of Caml and the object-oriented model of C\#, for .NET) \inlinelink{fsharp}{F\#}{http://research.microsoft.com/en-us/um/cambridge/projects/fsharp/}
provides an interface to .NET CLR's thread system.
Two  other levels  of abstraction  complete this  low level  layer for
concurrent programming: message-passing and asynchronous workflow.

In the same vein,
{\OCFMC}'s low  level threading mechanism allows to  build higher level
abstractions for  efficient and expressive  parallel programming.  For
instance,  the   Event  module  ({\OC}'s  module   to  offer  CML features)
is built upon that layer.
Following this, we can hope to use our work to allow existing parallel programming
systems built upon {\OC} to take advantage of multicore architectures.
We can cite CamlP3l \cite{mlws} \inlinelink{camlp3l}{OCamlP3l}{http://camlp3l.inria.fr/eng.htm}
 (skeleton programming: map, pipe, ...), 
Objective Caml-Flight \cite{CHAILLOUX:2003:HAL-00153378:1} \inlinelink{camlflight}{Objective Caml-Flight}{http://www-apr.lip6.fr/~chaillou/Public/Dev/nocf/}
 (data parallelism),
and
BSML \cite{HL2002:CMPPBOOK} \inlinelink{bsml}{BSML (Bulk Synchronous Parallel ML)}{http://frederic.loulergue.eu/research/bsmllib/bsml-0.4beta.html} (data parallelism and cost model). 



\section{Future works and conclusion} 


The experiment was successful since we effectively produced an
adaptation of {\OC} for current multicore architectures showing
promising performance results.

However, the strict guidelines we chose to follow, while founded, made
this project take a lot of time and restrained our possibilities.  For
instance, compiler modifications have been kept lightweight for we
hope they could be included as options in the standard distribution.

With less limitations, {\OCFMC} would probably lead to better results.  For instance, we have in mind some more intrusive
modifications to the compiler:
\begin{itemize}
\item Use of local heaps without backward pointers by exporting at
runtime the non-mutability property of constructed types, and thus
have local garbage collections without the need to stop everyone.
\item Insert checkpoints in the code so that the stop-the-world
mechanism could be more responsive and work even for non allocating
threads
\item Or dump more compiling information in the executable about the
state of the roots during execution to have less constraints for stopping the world. 
\end{itemize}


As a result, {\OCFMC} provides a runtime-level experiment platform to
develop new threading models or garbage collectors. It also brings the
possibility to design language-level concurrency abstractions over its
parallel shared memory low-level thread library.



\section*{Acknowledgements}

We are thankfull to Jane Street Capital for choosing {\OCFMC} as an Ocaml
Summer Project 2008 including the associated financial
participation.\\
We also thank Jon Harrop who tested an early version of OC4MC and gave
us his support during all the project. 








%



%

\insertlinks

\bibliographystyle{plain}
\bibliography{biboc4mc}


\end{document}